\documentclass[conference]{IEEEtran}
\IEEEoverridecommandlockouts
\usepackage[utf8x]{inputenc}
\usepackage[T1]{fontenc}
\usepackage{microtype,newtxtext,newtxmath} % use newpx[text|math] for palatino instead of times
\usepackage{mathtools,gensymb,eurosym,url,soul} % provides amsmath, generic symbols, URL management and strikeout
\usepackage{subcaption}% \usepackage[font=footnotesize,textfont=footnotesize]{subcaption}
\usepackage[colorinlistoftodos,prependcaption,textsize=tiny]{todonotes}
\usepackage{longtable,booktabs,array}
\usepackage[switch]{lineno}
\usepackage{acronym}
\usepackage{graphicx}
\usepackage{xcolor}%[usenames,dvipsnames,svgnames,table]{xcolor}
\usepackage{cite}
\usepackage{xr} % for referring to the paper document
\usepackage{subfiles} % subfiles
\usepackage{hyperref}

\hypersetup{colorlinks=true, breaklinks=true, %pagebackref=true,
  urlcolor=blue, linkcolor=blue,anchorcolor=blue,citecolor=blue,
  hypertexnames=true, final=true, 
  pdfpagemode = UseNone, %FullScreen, %UseThumbs, %UseOutlines,
  pdfauthor = {},
  pdftitle = {},   
  pdfsubject = {},
  pdfkeywords = {}
}

\graphicspath{{img/} {images/} {./}}
\def\BibTeX{{\rm B\kern-.05em{\sc i\kern-.025em b}\kern-.08em
		T\kern-.1667em\lower.7ex\hbox{E}\kern-.125emX}}

\begin{document}
\acrodef{ADC}[ADC]{Analog to Digital Converter}
\acrodef{ADEXP}[AdExp-I\&F]{Adaptive-Exponential Integrate and Fire}
\acrodef{ADM}[ADM]{Asynchronous Delta Modulator}
\acrodef{AER}[AER]{Address-Event Representation}
\acrodef{AEX}[AEX]{AER EXtension board}
\acrodef{AE}[AE]{Address-Event}
\acrodef{AFM}[AFM]{Atomic Force Microscope}
\acrodef{AGC}[AGC]{Automatic Gain Control}
\acrodef{AMDA}[AMDA]{AER Motherboard with D/A converters}
\acrodef{ANN}[ANN]{Artificial Neural Network}
\acrodef{API}[API]{Application Programming Interface}
\acrodef{APMOM}[APMOM]{Alternate Polarity Metal On Metal}
\acrodef{ARM}[ARM]{Advanced RISC Machine}
\acrodef{ASIC}[ASIC]{Application Specific Integrated Circuit}
\acrodef{AdExp}[AdExp-IF]{Adaptive Exponential Integrate-and-Fire}
\acrodef{BCM}[BMC]{Bienenstock-Cooper-Munro}
\acrodef{BD}[BD]{Bundled Data}
\acrodef{BEOL}[BEOL]{Back-end of Line}
\acrodef{BG}[BG]{Bias Generator}
\acrodef{BMI}[BMI]{Brain-Machince Interface}
\acrodef{BTB}[BTB]{band-to-band tunnelling}
\acrodef{CAD}[CAD]{Computer Aided Design}
\acrodef{CAM}[CAM]{Content Addressable Memory}
\acrodef{CAVIAR}[CAVIAR]{Convolution AER Vision Architecture for Real-Time}
\acrodef{CA}[CA]{Cortical Automaton}
\acrodef{CCN}[CCN]{Cooperative and Competitive Network}
\acrodef{CDR}[CDR]{Clock-Data Recovery}
\acrodef{CFC}[CFC]{Current to Frequency Converter}
\acrodef{CHP}[CHP]{Communicating Hardware Processes}
\acrodef{CMIM}[CMIM]{Metal-insulator-metal Capacitor}
\acrodef{CML}[CML]{Current Mode Logic}
\acrodef{CMOL}[CMOL]{Hybrid CMOS nanoelectronic circuits}
\acrodef{CMOS}[CMOS]{Complementary Metal-Oxide-Semiconductor}
\acrodef{CNN}[CCN]{Convolutional Neural Network}
\acrodef{COTS}[COTS]{Commercial Off-The-Shelf}
\acrodef{CPG}[CPG]{Central Pattern Generator}
\acrodef{CPLD}[CPLD]{Complex Programmable Logic Device}
\acrodef{CPU}[CPU]{Central Processing Unit}
\acrodef{CSM}[CSM]{Cortical State Machine}
\acrodef{CSP}[CSP]{Constraint Satisfaction Problem}
\acrodef{CTXCTL}[CTXCTL]{CortexControl}
\acrodef{CV}[CV]{Coefficient of Variation}
\acrodef{DAC}[DAC]{Digital to Analog Converter}
\acrodef{DAS}[DAS]{Dynamic Auditory Sensor}
\acrodef{DAVIS}[DAVIS]{Dynamic and Active Pixel Vision Sensor}
\acrodef{DBN}[DBN]{Deep Belief Network}
\acrodef{DFA}[DFA]{Deterministic Finite Automaton}
\acrodef{DIBL}[DIBL]{drain-induced-barrier-lowering}
\acrodef{DI}[DI]{delay insensitive}
\acrodef{DMA}[DMA]{Direct Memory Access}
\acrodef{DNF}[DNF]{Dynamic Neural Field}
\acrodef{DNN}[DNN]{Deep Neural Network}
\acrodef{DOF}[DOF]{Degrees of Freedom}
\acrodef{DPE}[DPE]{Dynamic Parameter Estimation}
\acrodef{DPI}[DPI]{Differential Pair Integrator}
\acrodef{DRAM}[DRAM]{Dynamic Random Access Memory}
\acrodef{DRRZ}[DR-RZ]{Dual-Rail Return-to-Zero}
\acrodef{DR}[DR]{Dual Rail}
\acrodef{DSP}[DSP]{Digital Signal Processor}
\acrodef{DVS}[DVS]{Dynamic Vision Sensor}
\acrodef{DYNAP}[DYNAP]{Dynamic Neuromorphic Asynchronous Processor}
\acrodef{EBL}[EBL]{Electron Beam Lithography}
\acrodef{EDVAC}[EDVAC]{Electronic Discrete Variable Automatic Computer}
\acrodef{EEG}[EEG]{electroencephalography}
\acrodef{EIN}[EIN]{Excitatory-Inhibitory Network}
\acrodef{EM}[EM]{Expectation Maximization}
\acrodef{EPSC}[EPSC]{Excitatory Post-Synaptic Current}
\acrodef{EPSP}[EPSP]{Excitatory Post-Synaptic Potential}
\acrodef{EZ}[EZ]{Epileptogenic Zone}
\acrodef{FDSOI}[FDSOI]{Fully-Depleted Silicon on Insulator}
\acrodef{FET}[FET]{Field-Effect Transistor}
\acrodef{FFT}[FFT]{Fast Fourier Transform}
\acrodef{FI}[F-I]{Frequency-Current}
\acrodef{FPGA}[FPGA]{Field Programmable Gate Array}
\acrodef{FR}[FR]{Fast Ripple}
\acrodef{FSA}[FSA]{Finite State Automaton}
\acrodef{FSM}[FSM]{Finite State Machine}
\acrodef{GIDL}[GIDL]{gate-induced-drain-leakage}
\acrodef{GOPS}[GOPS]{Giga-Operations per Second}
\acrodef{GPU}[GPU]{Graphical Processing Unit}
\acrodef{GUI}[GUI]{Graphical User Interface}
\acrodef{HAL}[HAL]{Hardware Abstraction Layer}
\acrodef{HFO}[HFO]{High Frequency Oscillation}
\acrodef{HH}[H\&H]{Hodgkin \& Huxley}
\acrodef{HMM}[HMM]{Hidden Markov Model}
\acrodef{HRS}[HRS]{High-Resistive State}
\acrodef{HR}[HR]{Human Readable}
\acrodef{HSE}[HSE]{Handshaking Expansion}
\acrodef{HW}[HW]{Hardware}
\acrodef{ICT}[ICT]{Information and Communication Technology}
\acrodef{IC}[IC]{Integrated Circuit}
\acrodef{IEEG}[iEEG]{intracranial electroencephalography}
\acrodef{IF2DWTA}[IF2DWTA]{Integrate \& Fire 2--Dimensional WTA}
\acrodef{IFSLWTA}[IFSLWTA]{Integrate \& Fire Stop Learning WTA}
\acrodef{IF}[I\&F]{Integrate-and-Fire}
\acrodef{IMU}[IMU]{Inertial Measurement Unit}
\acrodef{INCF}[INCF]{International Neuroinformatics Coordinating Facility}
\acrodef{INI}[INI]{Institute of Neuroinformatics}
\acrodef{IO}[I/O]{Input/Output}
\acrodef{IPSC}[IPSC]{Inhibitory Post-Synaptic Current}
\acrodef{IPSP}[IPSP]{Inhibitory Post-Synaptic Potential}
\acrodef{IP}[IP]{Intellectual Property}
\acrodef{ISI}[ISI]{Inter-Spike Interval}
\acrodef{IoT}[IoT]{Internet of Things}
\acrodef{JFLAP}[JFLAP]{Java - Formal Languages and Automata Package}
\acrodef{LEDR}[LEDR]{Level-Encoded Dual-Rail}
\acrodef{LFP}[LFP]{Local Field Potential}
\acrodef{LLC}[LLC]{Low Leakage Cell}
\acrodef{LNA}[LNA]{Low-Noise Amplifier}
\acrodef{LPF}[LPF]{Low Pass Filter}
\acrodef{LRS}[LRS]{Low-Resistive State}
\acrodef{LSM}[LSM]{Liquid State Machine}
\acrodef{LTD}[LTD]{Long Term Depression}
\acrodef{LTI}[LTI]{Linear Time-Invariant}
\acrodef{LTP}[LTP]{Long Term Potentiation}
\acrodef{LTU}[LTU]{Linear Threshold Unit}
\acrodef{LUT}[LUT]{Look-Up Table}
\acrodef{LVDS}[LVDS]{Low Voltage Differential Signaling}
\acrodef{MCMC}[MCMC]{Markov-Chain Monte Carlo}
\acrodef{MEMS}[MEMS]{Micro Electro Mechanical System}
\acrodef{MFR}[MFR]{Mean Firing Rate}
\acrodef{MIM}[MIM]{Metal Insulator Metal}
\acrodef{MLP}[MLP]{Multilayer Perceptron}
\acrodef{MOSCAP}[MOSCAP]{Metal Oxide Semiconductor Capacitor}
\acrodef{MOSFET}[MOSFET]{Metal Oxide Semiconductor Field-Effect Transistor}
\acrodef{MOS}[MOS]{Metal Oxide Semiconductor}
\acrodef{MRI}[MRI]{Magnetic Resonance Imaging}
\acrodef{NDFSM}[NDFSM]{Non-deterministic Finite State Machine} 
\acrodef{ND}[ND]{Noise-Driven}
\acrodef{NEF}[NEF]{Neural Engineering Framework}
\acrodef{NHML}[NHML]{Neuromorphic Hardware Mark-up Language}
\acrodef{NIL}[NIL]{Nano-Imprint Lithography}
\acrodef{NMDA}[NMDA]{N-Methyl-D-Aspartate}
\acrodef{NME}[NE]{Neuromorphic Engineering}
\acrodef{NN}[NN]{Neural Network}
\acrodef{NRZ}[NRZ]{Non-Return-to-Zero}
\acrodef{NSM}[NSM]{Neural State Machine}
\acrodef{OR}[OR]{Operating Room}
\acrodef{OTA}[OTA]{Operational Transconductance Amplifier}
\acrodef{PCB}[PCB]{Printed Circuit Board}
\acrodef{PCHB}[PCHB]{Pre-Charge Half-Buffer}
\acrodef{PCM}[PCM]{Phase Change Memory}
\acrodef{PE}[PE]{Phase Encoding}
\acrodef{PFA}[PFA]{Probabilistic Finite Automaton}
\acrodef{PFC}[PFC]{prefrontal cortex}
\acrodef{PFM}[PFM]{Pulse Frequency Modulation}
\acrodef{PR}[PR]{Production Rule}
\acrodef{PSC}[PSC]{Post-Synaptic Current}
\acrodef{PSP}[PSP]{Post-Synaptic Potential}
\acrodef{PSTH}[PSTH]{Peri-Stimulus Time Histogram}
\acrodef{QDI}[QDI]{Quasi Delay Insensitive}
\acrodef{RAM}[RAM]{Random Access Memory}
\acrodef{RA}[RA]{Resected Area}
\acrodef{RDF}[RDF]{random dopant fluctuation}
\acrodef{RELU}[ReLu]{Rectified Linear Unit}
\acrodef{RLS}[RLS]{Recursive Least-Squares}
\acrodef{RMSE}[RMSE]{Root Mean Squared-Error}
\acrodef{RMS}[RMS]{Root Mean Squared}
\acrodef{RNN}[RNN]{Recurrent Neural Networks}
\acrodef{ROLLS}[ROLLS]{Reconfigurable On-Line Learning Spiking}
\acrodef{RRAM}[R-RAM]{Resistive Random Access Memory}
\acrodef{R}[R]{Ripples}
\acrodef{SAC}[SAC]{Selective Attention Chip}
\acrodef{SAT}[SAT]{Boolean Satisfiability Problem}
\acrodef{SCX}[SCX]{Silicon CorteX}
\acrodef{SD}[SD]{Signal-Driven}
\acrodef{SEM}[SEM]{Spike-based Expectation Maximization}
\acrodef{SLAM}[SLAM]{Simultaneous Localization and Mapping}
\acrodef{SNN}[SNN]{Spiking Neural Network}
\acrodef{SNR}[SNR]{Signal to Noise Ratio}
\acrodef{SOC}[SOC]{System-On-Chip}
\acrodef{SOI}[SOI]{Silicon on Insulator}
\acrodef{SOZ}[SOZ]{Seizure Onset Zone}
\acrodef{SP}[SP]{Separation Property}
\acrodef{SRAM}[SRAM]{Static Random Access Memory}
\acrodef{STDP}[STDP]{Spike-Timing Dependent Plasticity}
\acrodef{STD}[STD]{Short-Term Depression}
\acrodef{STP}[STP]{Short-Term Plasticity}
\acrodef{STT-MRAM}[STT-MRAM]{Spin-Transfer Torque Magnetic Random Access Memory}
\acrodef{STT}[STT]{Spin-Transfer Torque}
\acrodef{SW}[SW]{Software}
\acrodef{TCAM}[TCAM]{Ternary Content-Addressable Memory}
\acrodef{TFT}[TFT]{Thin Film Transistor}
\acrodef{TLE}[TLE]{Temporal Lobe Epilepsy}
\acrodef{USB}[USB]{Universal Serial Bus}
\acrodef{VHDL}[VHDL]{VHSIC Hardware Description Language}
\acrodef{VLSI}[VLSI]{Very Large Scale Integration}
\acrodef{VOR}[VOR]{Vestibulo-Ocular Reflex}
\acrodef{WCST}[WCST]{Wisconsin Card Sorting Test}
\acrodef{WTA}[WTA]{Winner-Take-All}
\acrodef{XML}[XML]{eXtensible Mark-up Language}
\acrodef{divmod3}[DIVMOD3]{divisibility of a number by three}
\acrodef{hWTA}[hWTA]{hard Winner-Take-All}
\acrodef{sWTA}[sWTA]{soft Winner-Take-All}

\title{FPGA Implementation of An Event-driven Saliency-based Selective Attention Model
\thanks{This work was supported by the European Research Council (ERC) under the European Union’s Horizon 2020 Research and Innovation Program Grant Agreement No. 724295 (NeuroAgents).}
}
% \title{Real Time Event-driven Foveal Vision} 

% \markboth{IEEE TRANSACTION ON CIRCUITS AND SYSTEMS II: EXPRESS BRIEFS}%
% {Shell \MakeLowercase{\textit{et al.}}: Bare Demo of IEEEtran.cls for Journals}

\author{\IEEEauthorblockN{ Hajar Asgari}
	\IEEEauthorblockA{\textit{Institute of Neuroinformatics} \\
		\textit{ University of Zurich and ETH Zurich}\\
		Zurich, Switzerland \\
		hajar@ini.uzh.ch}
	\and
	\IEEEauthorblockN{ Nicoletta Risi}
	\IEEEauthorblockA{\textit{Bio-Inspired Circuits and Systems
Lab} \\
		\textit{University of Groningen}\\
		Groningen, Netherlands \\
		n.risi@rug.nl}
	\and
	\IEEEauthorblockN{ Giacomo Indiveri}
	\IEEEauthorblockA{\textit{Institute of Neuroinformatics} \\
	\textit{ University of Zurich and ETH Zurich}\\
	Zurich, Switzerland \\
	giacomo@ini.uzh.ch}
}

\maketitle
\begin{abstract}
Artificial vision systems of autonomous agents face very difficult challenges, as their vision sensors are required to transmit vast amounts of information to the processing stages, and to process it in real time.
One first approach to reduce the data transmission is to use event-based vision sensors, whose pixels produce events only when there are changes in the input. However, even for event-based vision, transmission and processing of visual data can be quite onerous. Currently these challenges are solved by using high-speed communication links and powerful machine vision processing hardware. But if resources are limited, instead of processing all the sensory information in parallel, an effective strategy is to divide the visual field into several small sub-regions, choose the region of highest saliency, process it and shift serially the focus of attention to regions of decreasing saliency. This strategy, commonly used also by the visual system of many animals, is typically referred to as ``selective attention''. Here we present a digital architecture implementing a saliency-based selective visual attention model for processing asynchronous event-based sensory information received from a \ac{DVS}. 
For ease of prototyping, we use a standard digital design flow and map the architecture on a \ac{FPGA}. 
We describe the architecture block diagram highlighting the efficient use of the available hardware resources demonstrated through experimental results exploiting a hardware setup where the \ac{FPGA} interfaced with the \ac{DVS} camera.

%  Artificial vision systems of autonomous agents are required to process vast amounts of information in real time. Event-based vision sensors greatly reduce the amount of data produced, however, even then, fast and efficient vision remains an open challenge. An effective strategy for solving this problem is to divide the visual field into small sub-regions and process them sequentially, starting from the ones of highest saliency. This strategy is typically referred to as ``selective attention''. Here we present a digital architecture implementing a saliency-based selective visual attention model for processing event-based data received from a DVS. We describe the architecture block diagram and demonstrate experimental results measured from the FPGA interfaced to the DVS. 
\end{abstract}


\IEEEpeerreviewmaketitle

\section{Introduction} 

Event-based vision sensors are becoming increasingly popular for a wide range of vision processing applications that require low latency and low power consumption~\cite{Gallego_etal20}.
As these constraints are the same ones faced by biological vision systems, it can be useful to study the computational strategies used by animal vision systems to optimize the performance of artificial event-based ones.
In particular, a prominent strategy that biology uses to carry out efficient vision processing with limited resources is to use \emph{selective attention}~\cite{Rizzolatti83, Koch_Ullman85,Itti_Koch01}.
Instead of simultaneously processing all the information that can be provided by the input sensors, this strategy serially selects \emph{salient} sub-regions of the input space shifting from sub-region to sub-region, and processing them sequentially.
These sub-regions are selected by choosing the region of the visual space that has the highest value in the corresponding ``saliency map''. 
In Itti's bottom-up selective attention model~\cite{Itti_Koch01}, saliency maps are computed by combining multiple visual features, such as color contrast, texture, motion, etc.~\cite{Itti_Koch01}.

As one of the most relevant features in these saliency maps is given by motion cues and dynamic changes in the visual scene (e.g., produced by stationary but flashing LEDs), here we focus on the activity of the \ac{DVS} sensor~\cite{DVS}, which naturally correlates with these salient features~\cite{Thompson12}.
The use of this sensor reduces the amount of data transferred to the processing areas and, consequently, the amount of power required to process it. 
Based on this, we present an event-based visual processing system that uses a biologically plausible computational model of a saliency-based form of bottom-up attention, to select and sequentially process sub-regions of the visual field.
While early attempts to implement selective attention models in neuromorphic hardware have focused on mixed-signal analog/digital implementations~\cite{Horiuchi_etal97,Morris_etal98,Indiveri00a,Indiveri01a,Bartolozzi_Indiveri09a}, few have been done using a standard digital design flow~\cite{Kim_etal11b,Molin_etal21}. Here we implement a saliency-based event-driven attention mechanism at pixel level which depend on the exponential decay of the pixel states. We use digital \ac{FPGA} devices for fast prototyping, while still allowing real-time processing of events streamed from the \ac{DVS}.
The active-vision system that we present carries out computation in parallel using an event-based asynchronous communication infrastructure that employs the \ac{AER}~\cite{Lazzaro_etal93, Boahen99, Deiss_etal98}. 

In Section~\ref{sec:background} we present the details of the selective attention model. Section~\ref{sec:event-based-saliency} describes the proposed event-based attention mechanism. In Section~\ref{sec:hardw-design-impl} we introduce the hardware architectures, and in Section~\ref{sec:results-discussion} we presents the simulation and experimental results of the proposed system. Finally, Section~\ref{sec:conclusion} concludes the paper.

\section{Background}
\label{sec:background}

	Selective attention is a powerful computational strategy used by biological systems to optimize the use of the limited computing resources and minimize latency and reaction times. It acts by selecting the most salient location of the visual scene, ignoring distractors and irrelevant information, and allocating the system's computational resources for fast and accurate processing of the information in that region. Subsequently, the system chooses other regions of the visual scene with decreasing saliency, processing them in a sequential manner. The choice of the regions to process can be influenced by ``bottom-up'' stimuli (e.g., such as high contrast or moving objects) or ``top-down'' preferences (such regions of the scene that ignore the sky, or that include faces).
	Several computational models of selective attention under both top-down and bottom-up influences have been proposed~\cite{Itti_etal98,Park_etal03,Borji_Itti13,Uejima_etal20}. 
	The first explicit, biologically plausible computational
	model of bottom-up saliency-based selective visual attention was proposed by Koch and Ullman in 1985~\cite{Koch_Ullman85}. 
	
	Inspired by biological vision systems, the first processing stage in any bottom-up attention model is computing different features such as intensity, contrast, color opponency, orientation, direction, or motion~\cite{Itti_Koch01}. 
	The different feature maps contribute with
	different weights to create a unique saliency map. The strength of the feature contributions can be influenced by top-down modulation or training~\cite{Itti_Koch01}.
        A \ac{WTA} network selects the most salient location in the saliency map and focuses the system's attention to it. As the region is being attended, an inhibition of return mechanism is used to suppress it, so that less salient regions can win the \ac{WTA} competition. Depending on the parameters governing its dynamics, this mechanism can allow the system to switch only between the two most salient regions, the top three, and so on.
        
        To test selective attention in real-time visual processing tasks, several event-driven saliency-based models have been proposed~\cite{Akolkar_etal15,Uejima_etal20}, as well as their hardware implementations, using neuromorphic circuits~\cite{Thakur_etal17,Galluppi_etal12,Bartolozzi_Indiveri09a}, and using \acp{FPGA}~\cite{Kim_etal11b,Molin_etal21}. However, almost all the \ac{FPGA} implementations made use of frame-based cameras~\cite{Molin_etal21}, and very few (only one to our knowledge) interfaced to silicon retina devices~\cite{Kim_etal11b}. In this paper we proposed a digital event-driven saliency-based attention model implemented on FPGA for prototyping to process event-streams of the DVS camera.

\section{Event-based Saliency Attention Model With Top-down Modulation}
\label{sec:event-based-saliency}
  
Here we present our \ac{FPGA}-compatible event-based selective attention architecture. 
As bottom-up selective attention models are especially appealing for hardware implementations, due to their modular and easily expandable nature~\cite{Indiveri01a}, we focus the bulk of our work on these sub-sets of models.
First, we describe how the model processes visual inputs from event-cameras and exploits the event-based nature of these cameras to compute saliency from  high-contrast changing visual stimuli; then we extend the bottom-up saliency model to incorporate a top-down modulation mechanism, which allows steering the focus of attention (FOA) according to the target task.

\subsection{Event-based Bottom-up saliency}
As opposed to the original computational model~\cite{Itti_etal98}, the system we designed uses only one feature channel that encodes the changes present in the visual scene, sensed by the \ac{DVS}.
Event-camera pixels, such as those present in the \ac{DVS}, operate independently and respond only to local changes in the light intensity, which makes them inherently act as local edge detectors of moving objects. 
In this respect, saliency is defined as the region of the visual scene that produces the highest number of events, which in turn corresponds to the area that has the highest number of high contrast moving objects.
%As a result, the asynchronous stream of events generated by event-cameras naturally extracts low-level features by encoding the scene contrast in the pixel spike timing.

Let the FOA be defined as a window of size $[m_x, m_y]$ centered around the most \textit{salient} pixel.
The \textit{event} $e(x,y,t,p)$ represents the brightness change detected by the pixel with coordinates \textit{(x,y)} at time \textit{t} and with polarity \textit{p} (with p=1 for contrast increases, on ON-event, and p=-1, or OFF-event, for contrast decreases). Then, the most \textit{salient} pixel at time \textit{t} can be defined as follows:
$$ P^*(t) = \max_{P} s_P(t) $$
with $s_P$ being the pixel state at time t. See~\ref{sect:SAL} for a detailed description of the algorithm, which updates the pixels state in an event-based manner.
	
	\subsection{Top-down Biasing}
	\label{method_tdb}
After receiving a pixel event and extracting information about the pixel's location, we can have a top-down bias ahead of the bottom-up processing. The top-down biasing can be executed in two methods: top-down gating and top-down modulation. In the ``top-down gating'' approach, all the areas out of a predefined region of interest are cut out. 
In the ``top-down modulation'' approach,  all the arriving pixels are processed but the changing gain of the state for each pixel is modulated according to the predefined conditions for the top-down biasing. %Based on the target application, we can apply different conditions to the top-down biasing mechanisms. For instance, we define some conditions on the pixel location to only take the events receiving from the upper half of the scene into account as the saliency location. When the system receives a stream of events, they are processed with a combination of top-down biasing mechanisms and bottom-up saliency-based algorithm. 

	%\subsection{Event-Based Selective Visual Attention}
\section{Hardware Design and Implementation}
\label{sec:hardw-design-impl}

\begin{figure}[h]
	\centering
	\includegraphics[width=1.0 \linewidth]{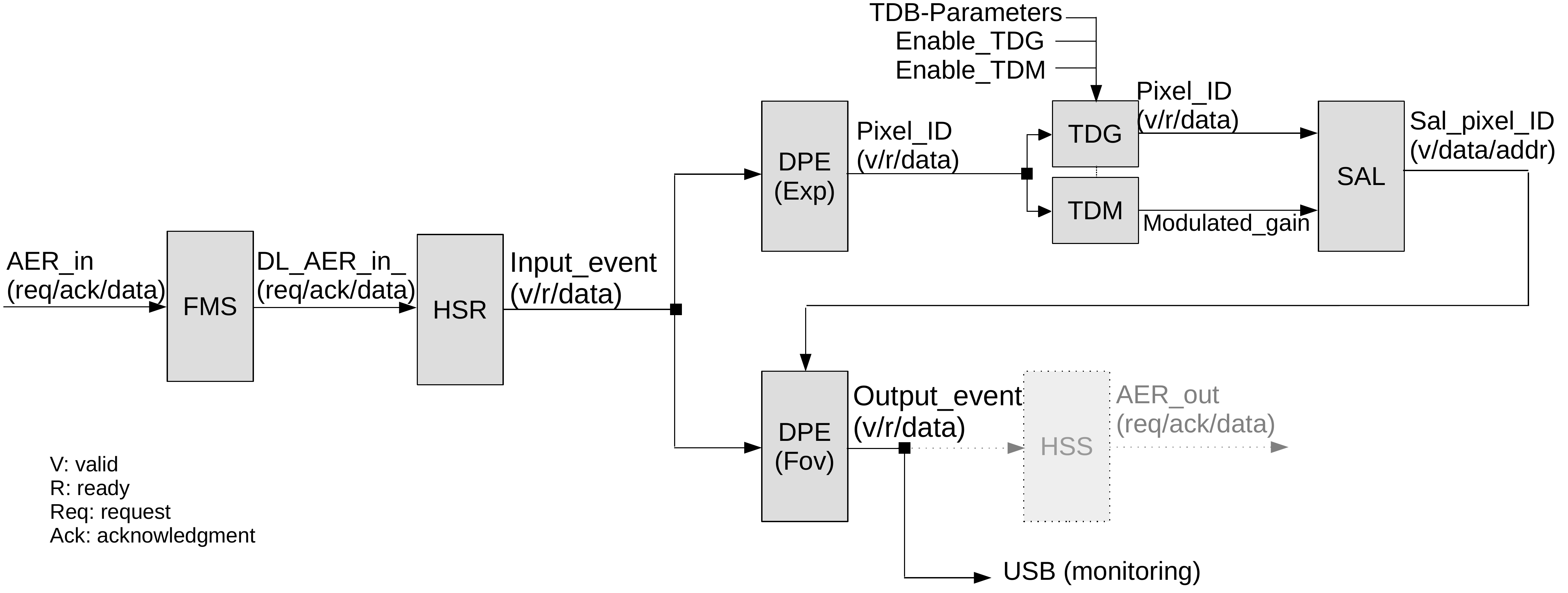}
	\caption{High-level block diagram of the network architecture. FMS: Flip-flop Meta-stability Synchronizer. HSR: Handshake Receiver. DPE (Exp): Data Processing Element in Exploring Mode. DPE (Fov): Data Processing Element in Fovea Mode. TDG: Top-Down Gating. TDM: Top-Down Modulation. SAL: Saliency block.  HSS (faded block): Handshake Sender.
	}
	\label{FOVEA}
\end{figure}

Fig.~\ref{FOVEA} shows a schematic of the proposed architecture, which consists of seven blocks.
The Handshake Receiver (HSR) handles the input events. HSR implements the standard 4-phase handshake AER protocol between the sender and the receiver block. The output events are monitored through a USB connection but can also be sent to the other event-based devices using a Handshake Sender (faded HSS block in Fig.~\ref{FOVEA}).
To prevent meta-stable states at the interface between different clock domains, a double flip-flop synchronizer (FMS) is applied before the input handshake. 
The core blocks of the architecture are described in more detail in the following sections.

\subsection{Data Processing Element(DPE)}
The event-camera included in our architecture is the DAVIS240C which employs a serial data format, i.e. \textit{x} and \textit{y} coordinates of the events are sent one after the other~\cite{DVS,Brandli_etal14}. 
Thus, the Data Processing Element block (or DPE) takes care of generating the 2D event coordinate by merging \textit{x} and \textit{y} input words and forwards the resulting pixel\_ID to the receiver block. Two instances of this block are used, which differ in the spatial resolution of the output events.

\begin{itemize}
\item
\texttt{DPE Exploring}, or \texttt{DPE (Exp)}, which operates on the full input resolution, extracts the pixel\_ID from the input events and forwards it to the Saliency block (SAL); SAL defines the most salient pixel according to the input events and a history of the previous events.
\item
\texttt{DPE Fovea}, or \texttt{DPE (Fov)}, which receives the coordinate of the most salient pixel by the SAL block and extracts only the events inside the FOA.
\end{itemize}
	
	\subsection{Top-Down Biasing Blocks} 	
	After exploring the input events and extracting the pixel\_ID, the top-down mechanism works on top of the bottom-up saliency. 
	In the proposed system, the top-down biasing mechanism is implemented using two different methods. Both are explained in more detail in \ref{method_tdb}.
	\subsubsection{Top-Down Gating (TDG)}
 This block receives the extracted pixel\_ID from DPE (Exp) block.
  When this block is enabled, via the input parameter \texttt{Enable\_TDG}, input events cross the TDG only if they meet the input top-down biasing conditions (\texttt{TDB\_parameters}). 
  
	\subsubsection{Top-Down Modulation (TDM)}
	TDM also receives the pixel\_IDs from the DPE (Exp). This block handles all the input events with a modulated state-changing gain. The amplitude of the modulation depends on the top-down biasing conditions. The highest modulation happens when the input events meet the TDM requirements.

\subsection{Saliency block (SAL)}
	\label{sect:SAL}
	\begin{figure}[h]
		\centering
		\includegraphics[width=1.0 \linewidth]{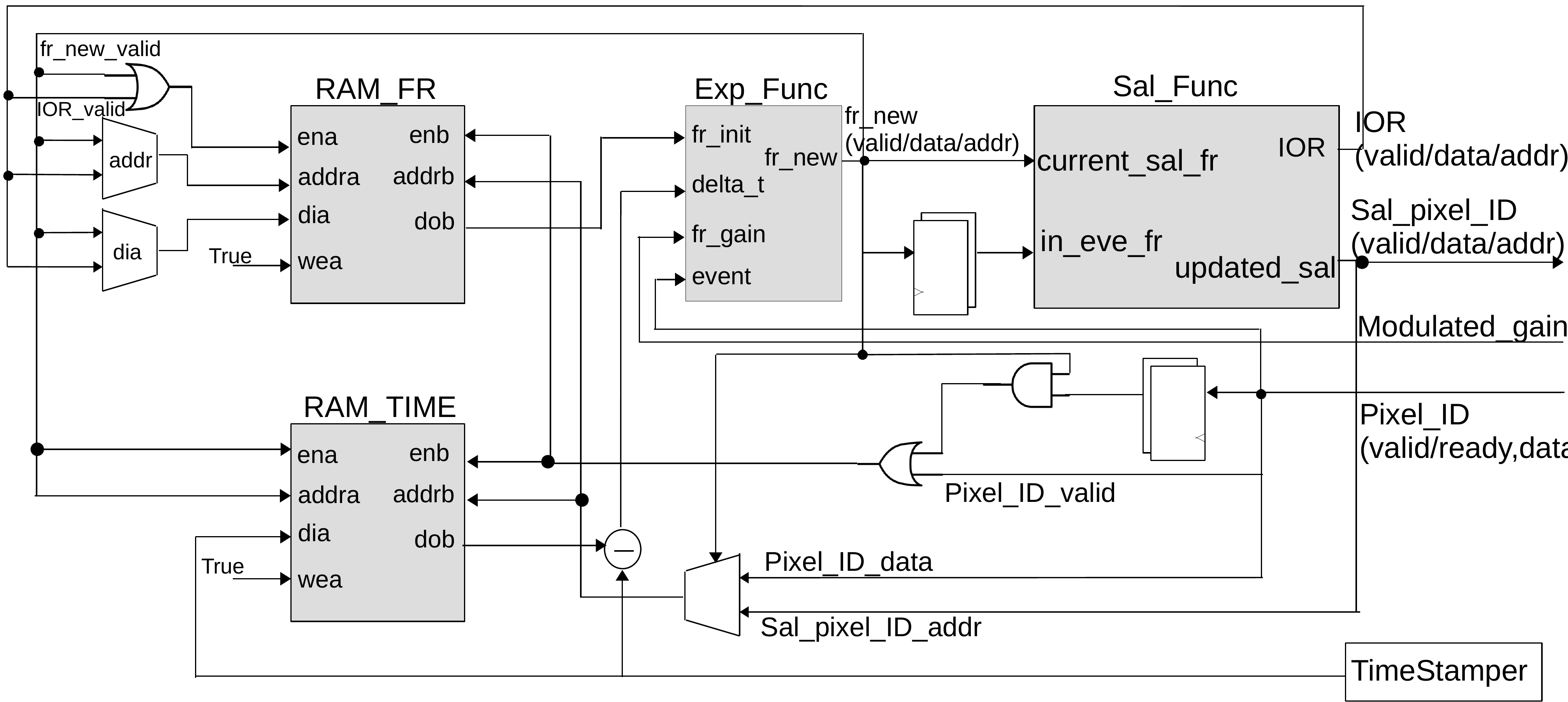}
		\caption{High-level block diagram of the saliency block (SAL). This block receives the input pixel\_ID and determines the most salient pixel.}
		\label{SAL}
	\end{figure}

 Fig.~\ref{SAL} illustrates the core blocks of the Saliency Block (or SAL) architecture. SAL  extracts the location of the most salient pixel. In the SAL architecture two separate Dual Port BRAM are used:
 	\texttt{RAM\_FR} stores the pixel state $s_P(t)$ as 21-bits registers (with fixed-point representation: s=1 bit for the sign, i=12 integer bits, and f=8 fractional bits);
 	\texttt{RAM\_TIME} stores timestamp of the last event generated at each pixel location.   
%\begin{itemize}
%\item
%\texttt{RAM\_FR}, which stores the pixel state $s_P(t)$ as 21-bits registers (with fixed-point representation: s=1 bit for the sign, i=12 integer bits, and f=8 fractional bits);
%\item
%\texttt{RAM\_TIME}, which stores timestamp of the last event generated at each pixel location.   
%\end{itemize}

Let $P^*$ be the current most salient pixel location with state $s^*$, which is selected based on all the previous input events.
At each incoming event from pixel $P$ at time point $t$, the SAL block performs the following steps:
\begin{enumerate}
\item
Read the last pixel event time stamp $t_{old}$ from the corresponding entry of the \texttt{RAM\_TIME}.
\item Read the latest pixel state $s_{P(t_{old})}$ at the last pixel event timestamp $t_{old}$ from the corresponding entry of the \texttt{RAM\_FR}.
\item Compute the updated pixel state $s_P(t)$ as follows:
\begin{equation}
s_P(t) = 1 + s_{P(t_{old})} \cdot \exp \left[(t_{old}-t)/\tau\right]
\end{equation}

with $\tau$ stored as input parameter. 
This operation is handled by \texttt{Exp\_Func} block (Fig.~\ref{SAL}), where the exponential decay term is implemented using the piece-wise linear approximation, to reduce the hardware resource utilization on the FPGA.
	
\item Store the new pixel state and timestamp in the corresponding BRAM entries.
\item Update the state of the current most salient pixel by repeating steps 1, 2, and 3 for $P^*$.
\item Compare the new pixel state to the state of the current most salient pixel $P^*$. If $s_P(t)>s*$: 
\begin{itemize}
\item
Update the most salient pixel location, which is forwarded to the DPE(Fov) block. 
\item
Excite the state of the updated most salient pixel as the center of FOV or $P^*$:
\begin{equation}
s[x,y] = s[x,y] + S_{plus}, \quad \forall (x,y) \in FOA
\end{equation}
 
\item
Inhibit the state of all pixels in the previous FOA:
\begin{equation}
s[x,y] = s[x,y] - S_{minus}, \quad \forall (x,y) \in FOA
\end{equation}
with $S_{plus}$ and $S_{minus}$ stored as input parameters for the highest excitation and inhibition values, respectively. 
\end{itemize}
By penalizing the pixels in the previously selected salient region, the last step adapts the Inhibition of Return (IOR) mechanism proposed in~\cite{Itti_Koch01} to our event-based implementation.
\end{enumerate}

\section{Results and Discussion}
\label{sec:results-discussion}

	The proposed system implements a digital electronic event-driven foveal vision.
	All parts of the design are described using the standard top-bottom digital design flow. 
	\\ 	
	As a proof of concept we implemented our electronic Foveal vision for processing event-camera inputs on a Kintex-7 XC7KT160T FPGA, which is hosted on the Opal Kelly XEM7360 board. 
	Based on the results from Table~\ref{comp:result}, the entire proposed network uses only less than 1 percent of the available Slice FFs and  LUTRAMs and DSPs, 2.31 percent of available Slice LUTs and 38 percent of available BRAMs. Furthermore, since the whole pipeline operates directly on the hardware, the system's latency is in the order of a few microseconds.  \\

\begin{figure}[h]
	\centering
	\includegraphics[width=1\columnwidth]{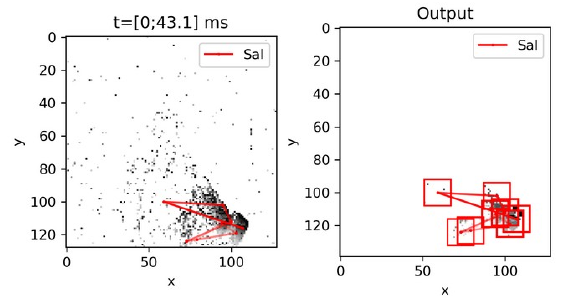}
	\caption{Time surface plots of the input (Left) and output (Right) of the system.} 
	\label{DVSGesture1}
\end{figure}

	\begin{table}[h]
		\renewcommand{\arraystretch}{1.1}
		\centering
		\caption{Resource utilization of the proposed digital electronic Foveal vision on Kintex-7(XC7KT160T).}
		\label{comp:result}
		\begin{tabular}{l l l l  }
			\hline
			& Proposed & Available & Utilization   \\ \hline
			Slice FFs   & 1646 & 202800& $0.81\%$\\ 
			Slice LUTs  & 3240 &  102400 & 2.31$\%$\\ 
			LUTRAMs & 33 & 35000 & 0.09 $\%$\\
			BRAMs & 123.5 & 325 & 38 $\%$\\
			DSPs    & 2 & 600 & 0.33 $\%$   \\
			%Max Frequency~(M.F)   &189&&\\
			Device  &Kintex.7  &&   \\		
			\hline
		\end{tabular}
	\end{table} 
To validate the architecture in more real-world scenarios, the design was tested with samples from the DVS128 Gesture Dataset~\cite{DVS128-Gesture-Dataset}, which contains DVS recordings from different hand gestures and illumination conditions.  Fig.~\ref{DVSGesture1} shows the trajectory of the most salient locations of a sample from the \textit{left hand clockwise} gesture. Fig.~\ref{DVSGesture1}-(Left) shows the time surface plot of the input events collected within a time window (t=[0;43.1] ms). The darker events are the most recent ones. The pixels shown in red are the most salient pixels selected across time. Fig.~\ref{DVSGesture1}-(Right) shows the output of the pipeline. Events are confined within a shifting $16\times16$ window, which corresponds to the most salient region at a given time point. 

\begin{figure}[h]
	\centering
	\includegraphics[width=0.8\columnwidth]{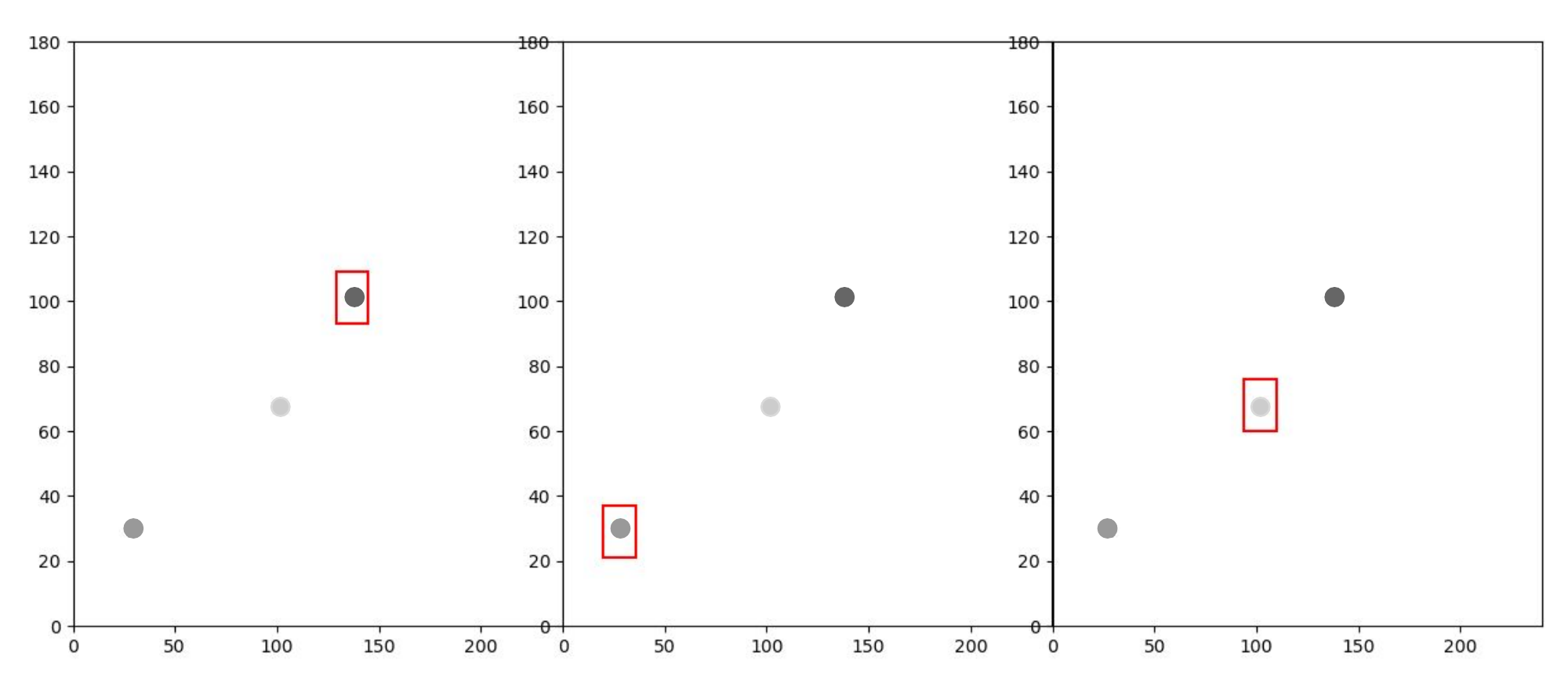}
	\caption{Examining the forgetting ability of the proposed system by applying a controlling input data stream. Darker colors show higher firing rates.} 
	\label{3point_plot}
\end{figure}

To examine the behavior of the proposed architecture in terms of the ability to attend to different salient locations, a controlled data stream was used to assess the system behavior. The data consists of 3 active pixels with different firing rates (depicted in gray scale levels in Fig.~\ref{3point_plot}). The architecture attends first to the pixel with the highest firing rate (dark gray point). Then it progressively shifts to the other active locations with decreasing firing rate (from gray to light gray regions). 
Thus, this result highlights how the proposed system can effectively shift the focus of attention towards different salient locations according to the local instantaneous firing rate.

\begin{figure}[h]
	\centering
	\includegraphics[width=1\columnwidth]{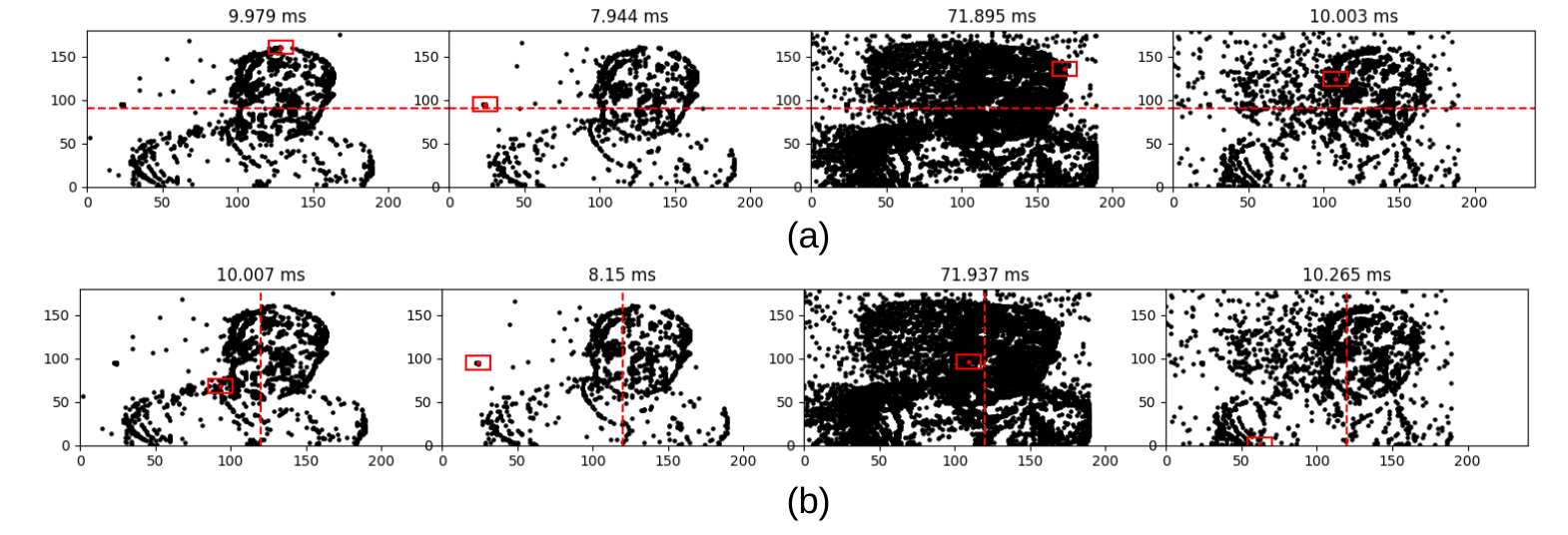}
	\caption{Bottom-up attention mechanism with two different top-down biasing parameters. The region of interest for finding the most salient location is: (a) upper-half. (b) left-half.} 
	\label{TDM_SIM_RESULT1}
\end{figure}

Fig.~\ref{TDM_SIM_RESULT1} illustrates the results with a top-down biasing condition. Exploring the events of half of the screen based on the application demands is an example of top-down biasing conditions. In this experiment, we have used input recordings from the DAVIS240. Unlike in the previous scenarios, here the pipeline is only attending to the most salient locations on the upper-half (Fig.~\ref{TDM_SIM_RESULT1}a) and left-half (Fig.~\ref{TDM_SIM_RESULT1}b) parts of the scene. 
The fovea window stays at one location for a limited time window (indicated on top of each plot), until the next most salient location is selected. 
 
\section{Conclusion}
\label{sec:conclusion}
We presented an implementation of a digital bio-inspired event-driven foveal vision on an FPGA. The system implements a simplified model of the bottom-up stimulus-driven saliency-based selective-attention mechanism.
The proposed network has been synthesized and as a proof of concept implemented on a Xilinx Kintext-7 FPGA device in a setup containing a DAVIS240 camera connected to a dedicated FPGA.
We demonstrated how the system can sequentially select the spatial locations of the most salient inputs present in the camera's field of view with experimental results. The system receives input signals in the form of address events, selects the spatial locations of the most salient inputs present in the camera’s field of view, and only transmits events from the selected sub-region.
This reduces the amount of data that needs to be processed, e.g., to extract features or detect specific targets, and enables the construction of a complex saliency-based vision processing system. In particular, it enables further neuromorphic event-based processing chips to process high-resolution images by using their limited resources (e.g., cores of 256 neurons~\cite{Merolla_etal14a, Moradi_etal18}) sequentially, as it happens in biology.

\bibliographystyle{ieeetr}
% \bibliography{biblio}
% \bibliography{references}
\bibliography{biblio/biblioncs}

\begin{thebibliography}{10}

\bibitem{Gallego_etal20}
G.~Gallego, T.~Delbr{\"u}ck, G.~Orchard, C.~Bartolozzi, B.~Taba, A.~Censi,
  S.~Leutenegger, A.~J. Davison, J.~Conradt, K.~Daniilidis, {\em et~al.},
  ``Event-based vision: A survey,'' {\em IEEE transactions on pattern analysis
  and machine intelligence}, vol.~44, no.~1, pp.~154--180, 2020.

\bibitem{Rizzolatti83}
G.~Rizzolatti, {\em Mechanisms of Selective Attention in Mammals},
  pp.~261--297.
\newblock Boston, MA: Springer US, 1983.

\bibitem{Koch_Ullman85}
C.~Koch and S.~Ullman, ``Shifts in selective visual-attention -- towards the
  underlying neural circuitry,'' {\em Human Neurobiology}, vol.~4, no.~4,
  pp.~219--227, 1985.

\bibitem{Itti_Koch01}
L.~Itti and C.~Koch, ``Computational modeling of visual attention,'' {\em
  Nature Reviews Neuroscience}, vol.~2, no.~3, pp.~194--203, 2001.

\bibitem{DVS}
``Product of inilabs: Dynamic vision sensor.''
  http://www.inilabs.com/products/dynamic-vision-sensors, 2009.

\bibitem{Thompson12}
J.~Thompson, ``Attention, biological motion, and action recognition,'' {\em
  Neuroimage}, vol.~59, no.~1, pp.~4--13, 2012.

\bibitem{Horiuchi_etal97}
T.~Horiuchi, T.~Morris, C.~Koch, and S.~DeWeerth, ``Analog {VLSI} circuits for
  attention-based, visual tracking,'' in {\em Advances in Neural Information
  Processing Systems ({NIPS})} (M.~Mozer, M.~Jordan, and T.~Petsche, eds.),
  vol.~9, pp.~706--712, MIT Press, 1997.

\bibitem{Morris_etal98}
T.~Morris, T.~Horiuchi, and S.~DeWeerth, ``Object-based selection within an
  analog {VLSI} visual attention system,'' {\em {IEEE} Transactions on Circuits
  and Systems {II}}, vol.~45, no.~12, pp.~1564--1572, 1998.

\bibitem{Indiveri00a}
G.~Indiveri, ``Modeling selective attention using a neuromorphic analog {VLSI}
  device,'' {\em Neural Computation}, vol.~12, no.~12, pp.~2857--2880, 2000.

\bibitem{Indiveri01a}
G.~Indiveri, ``A neuromorphic {VLSI} device for implementing {2-D} selective
  attention systems,'' {\em {IEEE} Transactions on Neural Networks}, vol.~12,
  pp.~1455--1463, Nov. 2001.

\bibitem{Bartolozzi_Indiveri09a}
C.~Bartolozzi and G.~Indiveri, ``Selective attention in multi-chip
  address-event systems,'' {\em Sensors}, vol.~9, no.~7, pp.~5076--5098, 2009.

\bibitem{Kim_etal11b}
B.~Kim, H.~Okuno, T.~Yagi, and M.~Lee, ``Implementation of visual attention
  system using artificial retina chip and bottom-up saliency map model,'' in
  {\em Neural Information Processing} (B.-L. Lu, L.~Zhang, and J.~Kwok, eds.),
  (Berlin, Heidelberg), pp.~416--423, Springer Berlin Heidelberg, 2011.

\bibitem{Molin_etal21}
J.~L. Molin, C.~S. Thakur, E.~Niebur, and R.~Etienne-Cummings, ``A neuromorphic
  proto-object based dynamic visual saliency model with a hybrid fpga
  implementation,'' {\em {IEEE} Transactions on Biomedical Circuits and
  Systems}, vol.~15, no.~3, pp.~580--594, 2021.

\bibitem{Lazzaro_etal93}
J.~Lazzaro, J.~Wawrzynek, M.~Mahowald, M.~Sivilotti, and D.~Gillespie,
  ``Silicon auditory processors as computer peripherals,'' {\em {IEEE}
  Transactions on Neural Networks}, vol.~4, pp.~523--528, 1993.

\bibitem{Boahen99}
K.~Boahen, ``Multiple pathways: Retinomorphic chips that see quadruple
  images,'' in {\em Proceedings of the Seventh International Conference on
  Microelectronics for Neural, Fuzzy and Bio-inspired Systems; Microneuro'99},
  (Los Alamitos, CA), pp.~12--20, {IEEE} Computer Society, Apr. 1999.

\bibitem{Deiss_etal98}
S.~Deiss, R.~Douglas, and A.~Whatley, ``A pulse-coded {C}ommunications
  infrastructure for neuromorphic systems,'' in {\em Pulsed Neural Networks}
  (W.~Maass and C.~Bishop, eds.), ch.~6, pp.~157--78, MIT Press, 1998.

\bibitem{Itti_etal98}
L.~Itti, E.~Niebur, and C.~Koch, ``A model of saliency-based visual attention
  for rapid scene analysis,'' {\em {IEEE} Transactions on Pattern Analysis and
  Machine Intelligence}, vol.~20, no.~11, pp.~1254--1259, 1998.

\bibitem{Park_etal03}
S.~Park, S.~Ban, S.~J., L.~Minho, O.~Kaynak, E.~Alpaydin, E.~Oja, and X.~Lei,
  ``Implementation of visual attention system using bottom-up saliency map
  model,'' in {\em Lecture notes in computer science, Artificial neural
  networks and neural information processing - ICANN / ICONIP} (B.~Springer,
  ed.), vol.~2714, pp.~678--685, June 2003.

\bibitem{Borji_Itti13}
A.~Borji and L.~Itti, ``State-of-the-art in visual attention modeling,'' {\em
  {IEEE} Transactions on Pattern Analysis and Machine Intelligence}, vol.~35,
  no.~1, pp.~185--207, 2013.

\bibitem{Uejima_etal20}
T.~Uejima, E.~Niebur, and R.~Etienne-Cummings, ``Proto-object based saliency
  model with texture detection channel,'' {\em Frontiers in Computational
  Neuroscience}, vol.~14, 2020.

\bibitem{Akolkar_etal15}
H.~Akolkar, D.~R. Valeiras, R.~Benosman, and C.~Bartolozzi, ``Visual-auditory
  saliency detection using event-driven visual sensors,'' in {\em International
  Conference on Event-based Control, Communication, and Signal Processing
  ({EBCCSP})}, pp.~1--6, 2015.

\bibitem{Thakur_etal17}
C.~S. Thakur, J.~L. Molin, T.~Xiong, J.~Zhang, E.~Niebur, and
  R.~Etienne-Cummings, ``Neuromorphic visual saliency implementation using
  stochastic computation,'' in {\em International Symposium on Circuits and
  Systems, ({ISCAS})}, pp.~1--4, IEEE, May 2017.

\bibitem{Galluppi_etal12}
F.~Galluppi, K.~Brohan, S.~Davidson, T.~Serrano-Gotarredona, J.~Carrasco,
  B.~Linares-Barranco, and S.~Furber, ``A real-time, event-driven neuromorphic
  system for goal-directed attentional selection,'' in {\em Neural Information
  Processing}, pp.~226--233, Springer, 2012.

\bibitem{Brandli_etal14}
C.~Brandli, R.~Berner, M.~Yang, S.-C. Liu, and T.~Delbruck, ``A
  240{$\times$}180 130 d{B} 3 {$\mu$}s latency global shutter spatiotemporal
  vision sensor,'' {\em {IEEE} Journal of Solid-State Circuits}, vol.~49,
  no.~10, pp.~2333--2341, 2014.

\bibitem{DVS128-Gesture-Dataset}
``Dvs128 gesture dataset.'' IBM Website.

\bibitem{Merolla_etal14a}
P.~A. Merolla, J.~V. Arthur, R.~Alvarez-Icaza, A.~S. Cassidy, J.~Sawada,
  F.~Akopyan, B.~L. Jackson, N.~Imam, C.~Guo, Y.~Nakamura, B.~Brezzo, I.~Vo,
  S.~K. Esser, R.~Appuswamy, B.~Taba, A.~Amir, M.~D. Flickner, W.~P. Risk,
  R.~Manohar, and D.~S. Modha, ``A million spiking-neuron integrated circuit
  with a scalable communication network and interface,'' {\em Science},
  vol.~345, pp.~668--673, Aug. 2014.

\bibitem{Moradi_etal18}
S.~Moradi, N.~Qiao, F.~Stefanini, and G.~Indiveri, ``A scalable multicore
  architecture with heterogeneous memory structures for dynamic neuromorphic
  asynchronous processors ({DYNAPs}),'' {\em Biomedical Circuits and Systems,
  {IEEE} Transactions on}, vol.~12, pp.~106--122, Feb. 2018.

\end{thebibliography}
	
\end{document}